\documentclass[onecolumn,showpacs]{revtex4}

\usepackage{graphicx}
\usepackage{dcolumn}
\usepackage{bm}

\input epsf

\begin{document}

\title{\Large Quasi-spherical gravitational collapse and the role of initial data, anisotropy and inhomogeneity}

\author{\bf Subenoy Chakraborty}
\email{subenoyc@yahoo.co.in}
\author{\bf Ujjal Debnath}
\email{ujjaldebnath@yahoo.com}

 \affiliation{Department of
Mathematics, Jadavpur University, Calcutta-32, India.}

\date{\today}

\begin{abstract}
In this paper, the role of anisotropy and inhomogeneity has been
studied in quasi-spherical gravitational collapse. Also the role
of initial data has been investigated in characterizing the final
state of collapse. Finally, a linear transformation on the
initial data set has been presented and its impact has been
discussed.
\end{abstract}


\maketitle

\section{{\protect\normalsize \textbf{Introduction }}}

Gravitational collapse is an important and challenging issue in
Einstein gravity for the last two decades or more, particularly
after the formation of famous singularity theorems [1] and Cosmic
Censorship Conjecture (CCC) [2]. Moreover, the final outcome of
gravitational collapse [3] in the background of general relativity
is interesting both from the point of view of black hole physics
as well as its astrophysical implications. In fact, an extensive
study of gravitational collapse [4 - 9] has been carried out of
Tolman-Bondi-Lema\^{\i}tre (TBL) spherically symmetric space-times
containing irrotational dust to support or disprove the CCC. A
general conclusion from these studies is that a central curvature
singularity forms but its local or global visibility depends on
the initial data. Recently, mena et al [10] have introduced a
linear transformation on the initial data set which keeps $\rho$
and $\sigma$ to be invariant at any time instant except for the
initial ones and have discussed the consequences for the final
fate of collapse.\\

On the otherhand, there is very little progress in studying
non-spherical collapse due to the ambiguity of horizon formation
and the influence of gravitational radiation. However, in recent
past, an extensive study of irrotational dust collapse has been
done in quasi-spherical Szekeres space-time [11-13]. In this
paper, we have extended the work of Mena et al to quasi-spherical
Szekeres model and have investigated the effect of inhomogeneity
and anisotropy on the final nature of singularity. The paper is
organized as follows: a brief outline of the quasi-spherical dust
collapse has been presented in section II. Section III deals with
properties of inhomogeneity and anisotropy both initially and at
any time instant and then linear transformation on the initial
data set have been discussed. The paper ends with conclusion in
section IV.\\

\section{{\protect\normalsize \textbf{Quasi-spherical dust collapse}}}

The inhomogeneous quasi-spherical dust collapse in $n$-dimension
is represented by Szekeres' space-time with line element

\begin{equation}
ds^{2}=-dt^{2}+e^{2\alpha }dr^{2}+e^{2\beta
}\sum_{i=1}^{n-2}dx_{i}^{2}
\end{equation}%
where the metric coefficients $\alpha $ and $\beta $ are functions
of all the $n$ space-time co-ordinates and have explicit form
[12,14]

\begin{equation}
e^{\alpha}=\frac{R^{\prime}+R~\nu^{\prime}}{\sqrt{1+f(r)}}~,~~~~
e^{\beta}=R(t,r)~e^{\nu(r,x_{1},...,x_{n})}
\end{equation}

The evolution equation for $R$ is

\begin{equation}
\dot{R}^{2}=f(r)+\frac{F(r)}{R^{n-3}}
\end{equation}

and $\nu$ has the explicit form

\begin{equation}
e^{-\nu}=A(r)\sum_{i=1}^{n-2}x_{i}^{2}+\sum_{i=1}^{n-2}B_{i}(r)x_{i}+C(r)
\end{equation}

where $F(r)(>0)$ and $f(r)(>-1)$ are arbitrary functions of $r$
and $A(r),~B_{i}(r)$'s and $C(r)$ are arbitrary functions of $r$
alone with the restriction

\begin{equation}
\sum_{i=1}^{n-2}B_{i}^{2}-4AC=-1
\end{equation}

(dot and dash stands for partial differentiation with respect to $t$ and $r$ respectively).\\

The matter density and the anisotropy scalar are given by [12]

\begin{equation}
\rho(t,r,x_{1},...,x_{n-2}) =\frac{(n-2)}{2}\left[\frac{F^{\prime
}+(n-1)F\nu ^{\prime } }{R^{n-2}(R^{\prime }+R\nu ^{\prime
})}\right]
\end{equation}
and
\begin{equation}
\sigma(t,r,x_{1},...,x_{n-2})=\sqrt{\frac{n-2}{2(n-1)}}\left[\frac{R\dot{R}'-\dot{R}R'}{R(R'+R\nu')}\right]
\end{equation}

Now integrating the evolution equation (3) we have

\begin{equation}
t-t_{i}=\frac{2}{(n-1)\sqrt{F}}\left[r^{\frac{n-1}{2}}~_{2}F_{1}[\frac{1}{2},a,a+1,-\frac{f
r^{n-3}}{F}]-R^{\frac{n-1}{2}}~_{2}F_{1}[\frac{1}{2},a,a+1,-\frac{f
R^{n-3} }{F}]\right]
\end{equation}

where $R=r$ at the initial epoch $t=t_{i}$ and $_{2}F_{1}$ is the
usual hypergeometric function with $a=\frac{1}{2}+\frac{1}{n-3}$ .\\

If $t=t_{s}(r)$ stands for time of collapse of the $r$-th shell
i.e., $R(t_{s}(r),r)=0$ then we have

\begin{equation}
t_{s}(r)-t_{i}=\frac{2}{(n-1)\sqrt{F}}r^{\frac{n-1}{2}}~_{2}F_{1}[\frac{1}{2},a,a+1,-\frac{f
r^{n-3}}{F}]
\end{equation}

Further, at the time of formation of trapped surfaces (i.e.,
$t=t_{ah}(r)$) we have $R^{n-3}=F(r)$, so we get

\begin{equation}
t_{ah}(r)-t_{i}=\frac{2r^{\frac{n-1}{2}}}{(n-1)\sqrt{F}}~_{2}F_{1}[\frac{1}{2},a,a+1,-\frac{f
r^{n-3}}{F}]-\frac{2F^{\frac{1}{n-3}}}{n-1}~_{2}F_{1}[\frac{1}{2},a,a+1,-f]
\end{equation}

The time of formation of central singularity is given by

\begin{eqnarray}
t_{0}=t_{s}(0)=t_{i}+
\begin{array}{ll}lim~~~~\frac{2r^{\frac{n-1}{2}}}{(n-1)\sqrt{F}}~_{2}F_{1}[\frac{1}{2},a,a+1,-\frac{f
r^{n-3}}{F}]\\
r\rightarrow 0
\end{array}
\end{eqnarray}

For finite value of the above limit, we assume $F(r)$ and $f(r)$
to be in the following polynomial form near the central
singularity $(r=0)$

\begin{equation}\begin{array}{c}
F(r)=F_{0}r^{n-1}+F_{1}r^{n}+F_{2}r^{n+1}+.........,\\\\

f(r)=f_{0}r^{2}+f_{1}r^{3}+f_{2}r^{4}+.........
\end{array}
\end{equation}

Then equation (6) suggests that initial density profile
$\rho_{i}~(=\rho(t_{i},r))$ is smooth at the centre and we have

\begin{equation}
\rho_{i}(r)=\rho_{0}+\rho_{1}r+\rho_{2}r^{2}+.........
\end{equation}

Now using the initial condition $R(t_{i},r)=r$ the initial density
and shear have the expressions

\begin{equation}
\rho_{i}= \frac{(n-2)}{2}~\frac{F'+(n-1)F\nu'}{r^{n-2}(1+r\nu')}
\end{equation}
and

\begin{equation}
\sigma_{i}=\sqrt{\frac{n-2}{8(n-1)}}~\frac{\left[\{r F'-(n-1)F
\}+r^{n-3}(r f'-2f)\right]}{r^{\frac{n-1}{2}}\left(F+f r^{n-3}
\right)^{1/2}(1+r\nu')}
\end{equation}

Using the series expansions (12) and (13) in equation (14) and
comparing terms of equal powers in $r$ we have

\begin{equation}
\rho_{j}=\frac{(n+j-1)(n-2)}{2}F_{j},~~~j=0,1,2,...
\end{equation}

Further, the above series expansions result (after simplification)
an explicit expression for the time difference between the
formation of trapped surface and the occurrence of central
singularity as (upto leading order in $r$)

\begin{eqnarray*}
t_{ah}-t_{0}=-\frac{2}{n-1}F_{0}^{\frac{1}{n-3}}r^{\frac{n-1}{n-3}}-
\frac{r}{(n-1)\sqrt{F}}\left[\frac{F_{1}}{F_{0}}~_{2}F_{1}[\frac{1}{2},a,a+1,-\frac{f_{0}
}{F_{0}}]\right.
\end{eqnarray*}

\begin{equation}
\left.+\frac{(n-1)}{(3n-7)}\frac{f_{0}}{F_{0}}\left(\frac{f_{1}}{f_{0}}-\frac{F_{1}}{F_{0}}\right)
~_{2}F_{1}[\frac{3}{2},a+1,a+2,-\frac{f_{0} }{F_{0}}]\right]
\end{equation}

The necessary condition for existence of naked singularity is
characterized by $t_{ah}(r)\ge t_{0}$. Hence the nature is
completely specified by the initial data set ${\cal I}=\{F,f\}$.
It is to be noted that if the coefficient of $r$ in the
expression (17) vanishes then the end state of collapse also
depends on the dimension of the
space-time.\\

\section{{\protect\normalsize \textbf{A study on anisotropy and inhomogeneity }}}

In this section we shall study some characteristic of anisotropy
and inhomogeneity of the space-time, we shall state them in the
form of proposition.\\

\subsection{{\protect\normalsize \textbf{Some Results}}}

{\bf Proposition I}:  {\it If
$\sigma(t_{i},r)=0=\dot{\sigma}(t_{i},r)$ then the initial data
set is homogeneous and the initial space-time is homogeneous and
isotropic.}\\

{\it Proof} : The expression for $\dot{\sigma}(t_{i},r)$ is

\begin{eqnarray*}
\dot{\sigma}(t_{i},r,x_{1},...,x_{n-2})=\sqrt{\frac{n-2}{2(n-1)}}~\frac{(n-3)[(n-1)F-rF']}{2r(1+r\nu')}
\end{eqnarray*}

\begin{equation}
-\frac{\sigma_{i}\left[\frac{F(5-n+2r\nu')}{r^{n-3}}+f(1+r\nu')+r\left(f'+\frac{F'}{r^{n-3}}\right)
\right] }{2r(1+r\nu')\sqrt{\frac{F}{r^{n-3}}+f}}\hspace{-1.5in}
\end{equation}

Hence $\dot{\sigma}=0=\sigma_{i}$ implies
\begin{equation}
(n-1)F-rF'=0
\end{equation}
i.e.,
\begin{equation}
F=F_{0}r^{n-1}
\end{equation}
Also $\sigma_{i}=0$ results
\begin{equation}
\frac{rF'-(n-1)F}{r^{n-3}}+(rf'-2f)=0
\end{equation}
So using (19) we have
\begin{equation}
rf'-2f=0 ~~{\text i.e.},~~ f=f_{0}r^{2}.
\end{equation}

Thus the initial data is homogeneous.\\

The expression for initial density contrast is

\begin{eqnarray*}
\left.\frac{\partial\rho}{\partial
t}\right|_{t=t_{i}}=\frac{(n-2)}{2}\left[\frac{F''}{r^{n-2}(1+r\nu')}
-\frac{(n-2)F'}{r^{n-1}(1+r\nu')^{2}}\right.\hspace{1in}
\end{eqnarray*}

\begin{equation}
\left.
+\frac{\left\{-(n-1)(n-2)F\nu'+r\left(rF'-(n-1)F\right)\left((n-1)\nu'^{2}-\nu''\right)
\right\}}{r^{n-1}(1+r\nu')}
 \right]\hspace{-1in}
\end{equation}

which vanishes identically for the choice of $F$ in equation (20).
It is to be noted that this choice of $F$ simplifies the initial
density to

$$
\rho_{i}=\frac{(n-1)(n-2)}{2}~F_{0}
$$

which is homogeneous. Hence the initial space-time is also
homogeneous and isotropic.\\

{\bf Proposition II}:  {\it If $\sigma(t,r,x_{1},...,x_{n-2})=0$
then } {\it (i) the initial data set is homogeneous,} {\it (ii)
the space-time is homogeneous and isotropic,} {\it (iii) the end
state of collapse will always be a black
hole.}\\

{\it Proof} : From equation (7), $\sigma=0$ implies
$$
\frac{\dot{R}'}{R'}=\frac{\dot{R}}{R}
$$
which shows that $R=g(r)\mu (t)$. But at $t=t_{i}$, $R=r$, hence
we have
\begin{equation}
R(t,r)=r~\mu (t)
\end{equation}

Now substituting in the evolution equation for $R$ (i.e., eq.(3))
we have
\begin{equation}
\dot{\mu}^{2}=\frac{F_{0}}{\mu^{n-3}}+f_{0}
\end{equation}

with $F=F_{0}r^{n-1},~f=f_{0}r^{2}$ i.e., the initial data set is
homogeneous.\\

So the expression for density simplifies to
$$
\rho(t)=\frac{(n-1)(n-2)F_{0}}{2\mu^{n-1}}
$$
Hence the space-time is homogeneous and isotropic (by
condition).\\

Further, using this expression for $R$ (i.e., eq.(24)) the time
difference between the formation of trapped surfaces ($t_{ah}$)
and the occurrence of central singularity ($t_{0}$) becomes

\begin{equation}
t_{ah}-t_{0}=-\frac{2}{(n-1)}F_{0}^{\frac{1}{n-3}}r^{\frac{n-1}{n-3}}~_{2}F_{1}[\frac{1}{2},a,a+1,-f_{0}r^{2}]
\end{equation}

So always we have $t_{ah}<t_{0}$ and the end state of collapse
will be a black hole. It is to be noted that similar results holds
also for spherical collapse [14].\\

{\bf Proposition III}: {\it If $\rho_{i}'=0$ then the initial data
set is homogeneous as well as the initial space-time is
homogeneous and isotropic.}\\

{\it Proof} : The expression for $\rho'_{i}$ can be simplified to

\begin{eqnarray*}
\rho'_{i}=\frac{(n-2)}{2r^{n-1}(1+r\nu')^{2}}~\left[(1+r\nu')\{rF''-(n-2)F'
\}\right.
\end{eqnarray*}

\begin{equation}
\left.  +\{rF'-(n-1)F \} \{(n-2)+(n-1)r\nu'^{2}-r\nu'' \} \right]
\hspace{-1in}
\end{equation}

which shows that for $\rho'_{i}=0$ we can choose $F=F_{0}r^{n-1}$.
Also for regularity of initial shear at $r=0$ demands
$f=f_{0}r^{2}$. Hence the initial data set is homogeneous. For
this choice of $F$ and $f$, $\rho_{i}$ becomes constant and
$\sigma_{i}=0$. Therefore, the initial space-time is homogeneous
and isotropic.\\

{\bf Proposition IV}: {\it If the density contrast vanishes at any
time instant (i.e., $\rho'(t,r,x_{1},...,x_{n-2})=0$), starting
from regular initial data, then the space-time will be homogeneous
and isotropic and final state of collapse will be a black hole.}\\

{\it Proof} : Using the condition for regular initial data (i.e.,
$F=F_{0}r^{n-1},~ f=f_{0}r^{2}$) the expression for energy density
simplifies to

\begin{equation}
\rho(t,r,x_{1},...,x_{n-2})=\frac{(n-1)(n-2)}{2}~\frac{r^{n-2}(1+r\nu')F_{0}}{R^{n-2}(R'+R\nu')}
\end{equation}

which shows that if we choose $R=r\mu(t)$ then $\rho$ will be
independent of $r$ i.e., $\rho'(t,r,x_{1},...,x_{n-2})=0$. It is
to be noted that $\rho$ will also become independent of other
space co-ordinates. Also the choice $R=r\mu(t)$ results
$\sigma=0$. Hence the space-time is homogeneous and isotropic. In
proposition II, we have shown that $\sigma=0$ results to a black
hole as the final state of collapse. Hence the proof.  \\\

From the above propositions it is clear that if the space-time is
homogeneous and isotropic then the final outcome of collapsing
process will be definite (a black hole) but no such conclusion is
possible if the space-time is inhomogeneous and anisotropic.\\

\subsection{{\protect\normalsize \textbf{Transformation of the initial data set}}}

In the initial data set ${\cal I}=\{F,f\}$ if we make a linear
transformation [10]

\begin{equation}
{\cal L}_{T}~:~{\cal
I}~\rightarrow~\{a^{\frac{n-1}{2}}F,~af\}~~~~~(a>0,~~\text
{a~constant})
\end{equation}

then the scale factor $R$ changes as $R\rightarrow a^{1/2}R$ while
$(\rho,\sigma)$ are invariant under this transformation but not
the initial density and shear i.e., $(\rho_{i},\sigma_{i})$. In
other words, there are classes of values of $F$ and $f$ for which
we have the same values of $\rho$ and $\sigma$ but the initial
ones (namely $\rho_{i}$ and $\sigma_{i}$) are different for each
set of $(F,f)$. So it may happen that for one initial data set
$(F,f)$ we have naked singularity as the final end state but
choosing `$a$' appropriately we have the transformed $(F,f)$ for
which we may black hole as the final outcome of the collapsing
process but $\rho,~\sigma$ remain same for both the initial data
set. This has been shown numerically for spherical collapse by
Mena et al [10]. So we may conclude that energy density and
anisotropy scalar at any time instant can not uniquely
characterize the nature of the final state of collapse. But for
the initial nature of the space-time no such conclusion can be
made. In fact for a given initial inhomogeneity and anisotropy the
initial data set can be specified uniquely and hence final outcome
of the collapsing process may be characterized.\\

\section{{\protect\normalsize \textbf{Conclusions}}}

In this paper, we have studied properties of anisotropy and
inhomogeneity of the space-time both initially and at any epoch in
the context of gravitational collapse. We have obtained two
distinct features from these studies namely: (a) the matter
density and shear scalar at any instant remain invariant under the
linear transformation (29) and hence can not characterize the
nature of the final singularity, (b) the initial energy density
and shear scalar are not invariant under the linear transformation
(29) but together can uniquely specify the initial data set and
hence they may characterize the final outcome of the collapsing
process. This is a very peculiar feature of the physical process
because physical parameters like energy density, shear must have
consistent behaviour for all time, they should not have different
characters initially and at any epoch. It seems that the trouble
is with the initial choice of the scale factor. We shall discuss
the role of initial choice in a subsequent paper.\\

{\bf Acknowledgement:}\\

The authors are thankful to IUCAA for worm hospitality where the
major part of the work has been done. One of the authors (U.D)
thanks CSIR (Govt. of India) for the
award of a Senior Research Fellowship.\\

{\bf References:}\\
\\
$[1]$  S. W. Hawking and G. F. R. Ellis, The large scale structure
of space-time (Cambridge. Univ. Press,
Cambridge, England, 1973).\\
$[2]$  R. Penrose, {\it Riv. Nuovo Cimento} {\bf 1} 252 (1969); in
General Relativity, an Einstein
Centenary Volume, edited by S.W. Hawking and W. Israel (Camb. Univ. Press, Cambridge, 1979).\\
$[3]$  For recent reviews, see, e.g. P. S. Joshi, {\it Pramana}
{\bf 55} 529 (2000); C. Gundlach, {\it Living Rev. Rel.} {\bf 2} 4
(1999); A. Krolak, {\it Prog. Theo. Phys. Suppl.} {\bf 136} 45
(1999); R. Penrose, in Black holes and relativistic stars, ed. R.
M. Wald (Univ. of Chicago Press, 1998);
T. P. Singh, {\it gr-qc}/9805066.\\
$[4]$ P. S. Joshi and I. H. Dwivedi, \textit{Commun. Math. Phys.}
\textbf{166 } 117 (1994).\\
$[5]$ P. S. Joshi and I. H. Dwivedi, \textit{Class. Quantum Grav.}
\textbf{16 } 41 (1999).\\
$[6]$ K. Lake, \textit{Phys. Rev. Lett.} \textbf{68} 3129 (1992).\\
$[7]$ A. Ori and T. Piran, \textit{Phys. Rev. Lett.} \textbf{59} 2137 (1987).\\
$[8]$ T. Harada, \textit{Phys. Rev. D} \textbf{58} 104015
(1998).\\
$[9]$ P. S. Joshi, \textit{Global Aspects in Gravitation and
Cosmology, }(Oxford Univ. Press, Oxford, 1993).\\
$[10]$ F. C. Mena, B. C. Nolan and R. Tavakol, {\it
gr-qc}/0405041.\\
$[11]$ P. S. Joshi and A. Krolak, {\it Class. Quantum Grav.} {\bf 13} 3069 (1996).\\
$[12]$ U. Debnath, S. Chakraborty and J. D. Barrow, {\it Gen. Rel.
Grav.} {\bf 36} 231 (2004); U. Debnath and S. Chakraborty, {\it
JCAP}~ {\bf 05} 001 (2004).\\
$[13]$  P. Szekeres, {\it Phys. Rev. D} {\bf 12} 2941 (1975); S.
Chakraborty and U. Debnath, {\it gr-qc}/0304072; {\it Int. J. Mod. Phys. D} {\bf 13} 1085 (2004).\\
$[14]$ P.S.Joshi, N.Dadhich and R.Maartens, {\it Phys. Rev. D}
{\bf 65} 101501(R) (2002).\\

\end{document}